\documentclass[12pt,preprint]{aastex}

\shorttitle{Metal Enrichment}
\shortauthors{Cen \& Riquelme}

\newcommand{\msun}{M_{\odot}}
\begin{document}

\title{Lower Metal Enrichment of Virialized Gas in Minihalos}

\author{ Renyue Cen\altaffilmark{1} and Mario A. Riquelme\altaffilmark{2}}

\altaffiltext{1} {Princeton University Observatory,
		  Princeton University, Princeton, NJ 08544;
		  cen@astro.princeton.edu}
\altaffiltext{2} {Princeton University Observatory,
		  Princeton University, Princeton, NJ 08544;
		  marh@astro.princeton.edu}

\begin{abstract}
We differentiate between the metal enrichment of the gas in virialized minihalos 
and that of the intergalactic medium at high redshift,
pertinent to cosmological reionization,
with the initial expectation that gas in the high density 
regions within formed dark matter halos may be more robust
thus resistant to mixing with lower density intergalactic medium.
Using detailed hydrodynamic simulations of gas clouds in minihalos
subject to destructive processes associated with  
the encompassing intergalactic shocks carrying metal-enriched gas,
we find, as an example, that, for realistic shocks of velocities of $10-100$km/s,
more than ($90\%,65\%$) of the high density gas with $\rho \ge 500\rho_b$
inside a minihalo virialized at $z=10$ of mass $(10^7,10^6)\msun$
remains at a metallicity lower than 3\% of that of the intergalactic medium
by redshift $z=6$.
It may be expected that the high density gas in minihalos
becomes fuel for subsequent star formation, when they are 
incorporated into larger halos where efficient atomic cooling 
can induce gas condensation hence star formation.
Since minihalos virialize at high redshift when the universe
is not expected to have been significantly reionized,
the implication is that gas in virialized minihalos may provide 
an abundant reservoir of primordial gas to possibly allow for the 
formation of Population-III metal-free stars to extend to 
much lower redshift than otherwise expected based on
the enrichment of intergalactic medium.
\end{abstract}


\section{Introduction}

Recent observations of the high redshift ($z > 6$) quasar spectra from
the Sloan Digital Sky Survey (SDSS; \citealt{fan01,bec01,bar02,cen02})
and the cosmic microwave background fluctuations from the Wilkinson Microwave Anisotropy Probe
(WMAP; \citealt{spe06}) combine to paint a complicated and yet uncertain
reionization picture.
What is strongly suggested, however, is that 
the reionization process began at an early time, probably at $z\ge 15$.
If stars are primarily responsible for producing the ionizing photons,
the overall reionization picture would depend significantly
on how and when the transition from
metal-free Population III (Pop III) to metal-poor Population II (Pop
II) stars occurs (see, e.g., \citealt{omu00,bro01,sch02,mac03,sch03,bro03}),
especially if Pop-III starformation history is more extended than
usually thought \citep{wc07}.
This is based on
the finding that Pop III stars may be much more massive and hotter than Pop II stars
\citep{car84,lar98,abe00,her01,bro01,bkl01,nak01,bro02,omu03,mac03}
and hence efficient producers of ionizing photons of hydrogen
and helium.

The metallicity of the star-forming gas
plays several important roles in the physics of first stars.
First, the transition from Pop III to Pop II is
facilitated by the presence of a small amount of metals, in
particular, oxygen and carbon \citep{bro03}.
Thus, it is the amount of C and O, not
necessarily the total amount of ``metals", that determines the
transition \citep{fc04}. 
The yield patterns for (non-rotating) stars with mass in the range of $140-260\msun$
that explode via the pair-instability supernovae (PISN) 
and regular type II SNe are different.
In PISN case the
supernova ejecta is enriched by $\alpha$-elements, 
whereas the major products of SNII are hydrogen and helium with a small amount of heavy
elements (see, e.g., \citealt{woo95,heg02}).  
Consequently, the
transition from Pop III to Pop II stars may occur at different times,
depending on the IMF (e.g., \citealt{fc04}).

Second, while the ionizing photon production efficiency
depends only relatively weakly on the exact IMF, as long as the stars
are more massive than $\sim 10\msun$ (e.g., \citealt{tum04}),
its dependence on metallicity is strong, because
the effective temperature
of the stellar photosphere depends sensitively on the opacity hence metallicity 
of the stellar atmosphere.
The amount of metals produced depends on the IMF.
For example, in the most extreme case where all Pop-III stars
are more massive than, say, $\ge 270\msun$,
these stars may conclude by implosions to intermediate-mass 
black holes without giving out much metals to the surroundings.
However, exactly how massive Pop III stars are is uncertain. 
While simulations have suggested that Pop III stars may be more massive than
$100\,\msun$ (``very massive star'', VMS; \citealt{abe00,bro01}),
\citet{tan04} find that stellar feedback processes may
limit the mass of the Pop III stars to the range $30-100\,\msun$.  
Observationally, the VMS picture is advocated by
\cite{ohn01} and \cite{qia02}, based on an analysis of metal
yield patterns from pair-instability supernova (PISN) explosion of VMS
progenitors \citep{heg02}.
\cite{tum04}, \cite{dai04},
Umeda \& Nomoto (2003,2005)
and \cite{ven03},
on the other hand, 
argue that the general pattern in metal-poor halo stars,
in the Ly$\alpha$ forest and cosmic star formation history,
is more consistent with
the yield pattern of Type
II supernovae (SNII) perhaps with a lower cutoff of 10$\msun$.

Clearly, the metallicity of gas out of which stars are formed
is critically important.
The conventional picture that is often adopted goes as follows:
formed stars will eject metals into the IGM 
and eventually raise the metallicity of the IGM to above
the threshold for the Pop-III to Pop-II transition.
A somewhat refined version of this takes into account 
that the metallicity enrichment process of the IGM
is unlikely to be synchronous for different regions (e.g., \citealt{fl05}).
Here, we point out a possibly large difference
between the metallicity of the IGM and the metallicity
of the gas in minihalos. 
Since minihalos collapse at very high redshift (e.g., \citealt{wc07}),
the large amount of dense gas in minihalos thus may provide a primary fuel
for subsequent star formation, 
when eventually they are incorporated into 
large systems where efficient atomic cooling allows gas
to condense to form stars.

To quantify this possible difference between the metallicity 
of minihalo gas and that of the IGM, we study the 
stability and metal enrichment of minihalos subject to metal-rich shockwaves 
launched by supernovae explosions from large galaxies.
We will treat an idealized situation where a minihalo 
is subject to shock waves enriched with a chosen metallicity,
and we investigate how gas inside it may be contaminated by metals.
We will assume that there has been no star formation hence
no self-metal-enrichment in minihalos,
because of the lack of adequate coolants;
molecular hydrogen is assumed to have long been destroyed 
by Lyman-Werner photons produced by earlier stars elsewhere.
Because gas in minihalos is significantly overdense compared to
the IGM and is bounded by the gravitational potential wells 
produced by the dark matter halos,
mixing of metals into the gas in minihalos 
by metal-rich outflows from star-forming galaxies
should be expected to be different from that of the IGM.
As we will show,
the process of mixing
of metal-rich outflows with the gas in minihalos is quite incomplete.
Several authors 
%\citep{MurWhiBloLin93, M93 hereafter,KleMcKCol94,Din97,MinJonFerRyu97} 
(\citealt{MurWhiBloLin93}, M93 hereafter;
\citealt{KleMcKCol94, Din97, MinJonFerRyu97}) 
have addressed the problem of the stability of a 
non-self-gravitating gas cloud moving at the sound speed of the background medium, 
which is equivalent to a shockwave sweeping the gas cloud. They have found that the 
cloud gets disrupted after a time comparable to the dynamical time of the cloud. 
Here, we are interested in the self-gravitating case. In particular, we are interested in
minihalos that are gravitationally dominated by their dark matter content, and with
no cooling. 
A very similar case was already studied by M93, 
in the context of a two-phase medium, using 2-D simulations.
In this work we employ 3-d hydrodynamical simulations to study this problem. 
We simulated halos of mass $10^6\msun $ and $10^7\msun$ subject to 
shockwaves with velocities of
10, 30, 100, and 300 km/s. For the slowest cases of 10 and 30 km/s the halos are quite stable
and the gas inside the virial radius of the halos remains fairly 
uncontaminated after many dynamical times. Only for shock 
velocities of 100 and 300 km/s the halos start to be unstable, loosing significant 
fraction of their gas, and getting substantially enriched in their inner regions.

The paper is organized a follows. 
In \S 2 we specify the physical model for
the minihalos and shockwaves, and describe some
technical specification for the code we use. 
\S 3 presents our results,  followed by conclusions in \S 4.      

\section{Description of the model}\label{model}

We analyze the metal enrichment of gas in spherical minihalos with total 
virial masses of $10^6$ and $10^7 M_{\odot}$,
whose virial temperatures are $710$~K and $3295$~K, respectively,
at $z=10$.
Initially, the gas in minihalos is assumed to have
zero metallicity.
Then, the minihalo is exposed to an IGM sweeping through at a 
velocity of $V_s$ and metallicity $Z_{IGM}$,
and we quantify the evolution of the metallicity 
of the gas inside the minihalo.
We study four cases with $V_s = 10, 30, 100,$ and $300$ km/s
for each of the two choices of minihalo masses.

The gravitational potential of a halo is determined by their dark matter
and assumed not to change. 
The density of a virialized dark matter halo as a function of radius, $r$,
is given by the NFW \citep{NavFreWhi97} density profile:
\begin{eqnarray}
\rho_{DM}(r) = \frac{\rho_{crit}\delta_c}{u(1+u)^2},
\label{eq:nfw}
\end{eqnarray}
where $\rho_{crit}=3H(z)^2/8\pi G$ is the critical density of the universe at redshift $z$, $\delta_c = 200c^3/3\textrm{ }m(c)$, and $u=r/r_s$. The characteristic radius $r_s$ is defined in terms of the concentration parameter of the halo, $c$, that is a function of the halo mass and the redshift, and the virial radius, $r_{vir}$. The virial radius is defined in terms of the halo mass, $M_H$, by $(4\pi /3)r_{vir}^3 200 \rho_{crit}=M_H$, and the function $m(u) = \textrm{ln}(1+u) - u/(1+u)$. For the concentration parameter we adopt 
the fitting formula provided by \cite{Dol04}:
\begin{eqnarray}
c=\frac{9.59}{1+z}\big(\frac{M_H}{10^{14}h^{-1}M_{\odot}}\big)^{-0.102},
\label{eq:c}
\end{eqnarray}
based on computations of a $\Lambda$CDM cosmological model with $\Omega_m=0.3$, $\Omega_{\Lambda}=0.7$, $\Omega_b=0.045$, and $\sigma_8 = 0.9$.

Since the gravitational potential, $\phi$, is determined by the dark matter content of the minihalos, it will be given by:
\begin{eqnarray}
\phi(r) = \left\{
\begin{array}{ll}
-4\pi G \textrm{ }\delta_c \rho_{crit}\textrm{ }r_s^2 \frac{\textrm{ln}(1+u)}{u} & \textrm{for } u \leq d   \\
-4\pi G \textrm{ }\delta_c \rho_{crit}\textrm{ }r_s^2 \Big[\frac{\textrm{ln}(1+d)}{d} + m(d)\Big(\frac{1}{u}-\frac{1}{d}\Big)\Big] & \textrm{for } u > d,
\end{array}
\right.
\label{eq:potential}
\end{eqnarray}
where $r_sd$ is the radius at which the dark matter density of the halo equals the mean density of the universe.

For the gas we used a $X=0.76$, $Y=0.24$, and $Z=0$ composition. Its density and temperature profiles will be determined by assuming $P_g = \rho_g k_BT_g/(m_p\mu) \textrm{ } \alpha \textrm{ } \rho_g^{\gamma}$, where $\gamma$ is the polytropic index. Then we can write
\begin{eqnarray}
\rho_g(r) = \rho_c y(u) \textrm{  and  } T_g(r) = T_c y(u)^{\gamma - 1},
\label{eq:densandtemp}
\end{eqnarray} 
where $\rho_c$ and $T_c$ are the density and temperature at the center of the halo. Since the halos are in hydrodynamic equilibrium we find,
\begin{eqnarray}
y(u)^{\gamma - 1} = \left\{
\begin{array}{ll}
1 + \frac{\gamma-1}{\gamma}\frac{G M_H m_p\mu}{kT_c r_s m(c)} \Big(\frac{\textrm{ln}(1+u)}{u} - 1 \Big) & \textrm{for } u \leq d   \\
1 + \frac{\gamma-1}{\gamma}\frac{G M_H m_p\mu}{kT_c r_s m(c)} \Big(\frac{\textrm{ln}(1+d)}{d} - 1 + m(d)\Big(\frac{1}{u}-\frac{1}{d}\Big) \Big) & \textrm{for } u > d.
\end{array}
\right.
\label{eq:y}
\end{eqnarray}
We have three free parameters in our gas profile: $\rho_c$, $T_c$, and $\gamma$. The central density will be chosen such that at $r=r_{vir}$ the ratio between dark and baryonic matter densities is equal to $\Omega_m/\Omega_b$. The values for $T_c$ and $\gamma$ are determined considering that, according to many hydrodynamic simulations, in the outer part of the halos the gas density profile traces the dark matter density profile \citep{KomSel01}, i.e., 
\begin{eqnarray}
\frac{d\ln(\rho_{DM})}{d\ln(\rho_g)} = 1. 
\label{eq:trace}
\end{eqnarray} 
So $T_c$ and $\gamma$ were determined using the fitting formulae provided by \cite{KomSel01} that satisfy equation (\ref{eq:trace}) within a range $c/2 \lesssim u \lesssim 2c$:
\begin{eqnarray}
\begin{array}{lll}
T_c & = & \frac{G M_H m_p\mu}{3k r_s c}(0.00676(c-6.5)^2 + 0.206(c-6.5) + 2.48)   \\
\gamma & = & 1.15 + 0.01(c-6.5).
\end{array}
\label{eq:tcandgama}
\end{eqnarray} 
As mentioned above,
the minihalos will be shocked by a continuous, hot, metal-rich shockwave at a temperature $T_s=\frac{3V_s^2m_p}{16k_B}$, a mean density $\rho_s=4\bar{\rho}_b$, and a metallicity $Z_{IGM}$. We also introduce fluctuations in space and time in the density of the shockwave, i.e., 
\begin{eqnarray}
\rho_s = 4\rho_b(1+A\sin (\frac{2\pi y}{\lambda}+\phi_y)\sin (\frac{2\pi z}{\lambda}+\phi_z)\sin (\frac{2\pi t}{T}+\phi_t)),
\label{eq:shockdensity}
\end{eqnarray}
where $y$ and $z$ represent the two spatial coordinates perpendicular to $x$, the direction of propagation of the shockwave, $t$ is the time, $A$ and $\lambda$ represent the amplitude and the length of the fluctuation, and $\phi_i$ corresponds to an arbitrary phase in the coordinate $i$. The spatial phases were randomly chosen every time that $t/T$ became an integer number, where $T = \lambda/V_s$. 
What will be appropriate values for $A$ and $\lambda$?
At the redshifts of interest ($z\sim 6-10$), large atomic-cooling halos
start to become nonlinear, meaning that the density variance
is of order unity on the mass scales of $10^8-10^9\msun$, corresponding
length scales of $\sim 0.1$~Mpc.
By definition, when a certain mass scale $M$ become nonlinear, 
$A(M)\sim 1$.
We have experimented with values of $A=0.3-0.9$ 
and $\lambda$ $=$ 0.003, 0.01, and 0.03 Mpc/h in comoving units.  
Our results turned out to be nearly independent of the values for $A$ and $\lambda$
in the ranges of relevance.

Each simulation starts at $z=10$, when the IGM shockwave
enters the left face of our simulation cube.
We do not attempt to vary the background density with time,
aside from the variation imposed (see Equation (\ref{eq:shockdensity}) above).
We expect that, if the background density were allowed to decrease
with time, the metal enrichment of the gas in minihalos
may be reduced.
When there is a need to indicate a redshift during
the evolutionary phase of a minihalo,
we translate the elapsed time since $z=10$
to a certain redshift, 
using the standard cosmological model 
parameters \citep{spe06}.

%\section{The simulation}
%\label{sec:simulation}
We use the TVD hydrodynamics code \citep{cenetal03} to perform the described simulations.
The size of the boxes is chosen such that at the border of the box the gas density of the halo is equal to the mean baryonic density of the universe. So the comoving size of the boxes is 0.0191 and 0.0457 Mpc/h for $M_H = 10^6$ and $10^7 M_{\odot}$, respectively. 
For most of the simulations we use $256^3$ cells for each simulation. 
Our results seem to be convergent to a few percent accuracy 
as will be shown at the end of the next section.

\section{The results}
\label{sec:results}
In this section we analyze our results for the stability and chemical evolution from $z=10$ to $z=6$ for the two halos considered. 
Although for parameters
of the range considered
we observe different levels of instability and mixing, in all cases the gas in the inner region of the halos remains substantially less metallic than the IGM. 
For $M_H = 10^7M_{\odot}$ and $V_s = 10$ km/s almost all the mass at density higher than the virial density, $\rho_{vir}$ (= 49.3 $\rho_b$ for $M_H = 10^7M_{\odot}$) at $z=6$ 
has $Z < 0.03 Z_{IGM}$, whereas for $M_H = 10^6M_{\odot}$ and $V_s = 300$ km/s 
most of the mass at $\rho \ge \rho_{vir}$ (= 40.3 $\rho_b$ for $M_H = 10^6M_{\odot}$) remains with $Z < 0.3 Z_{IGM}$.

\begin{figure}[htp]
\centering
\includegraphics[width=0.43\textwidth,angle=90]{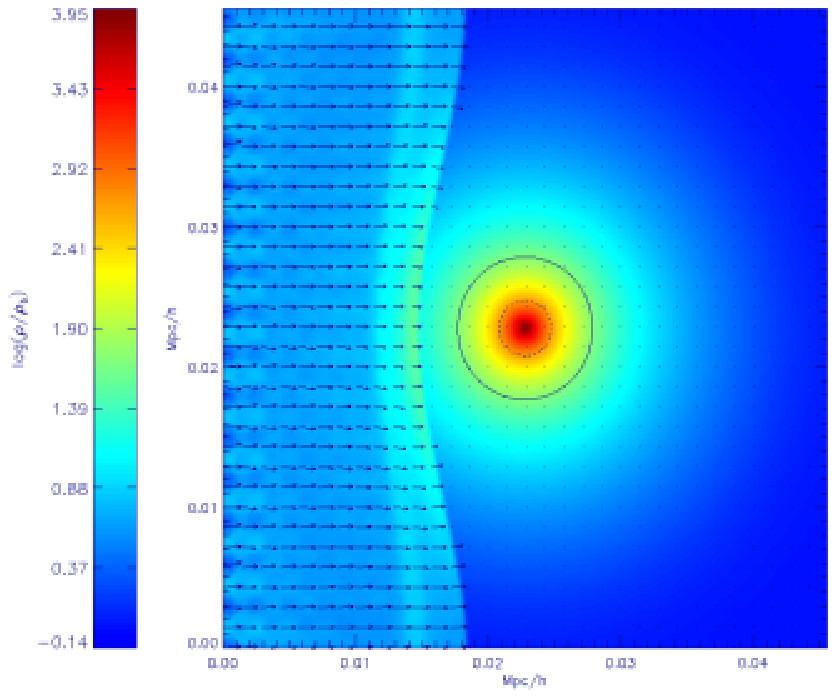}
%\hfill
%\includegraphics[width=0.43\textwidth,angle=90]{d_v30m7z6.ps}\\
%\includegraphics[width=0.43\textwidth,angle=90]{m_v30m7z9.ps}
\includegraphics[width=0.43\textwidth,angle=90]{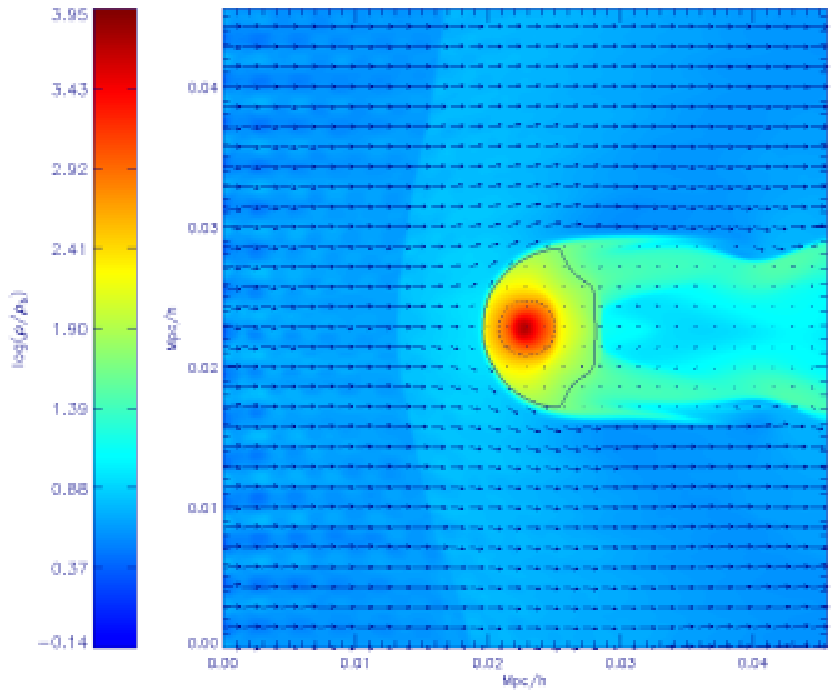}\\
\includegraphics[width=0.43\textwidth,angle=90]{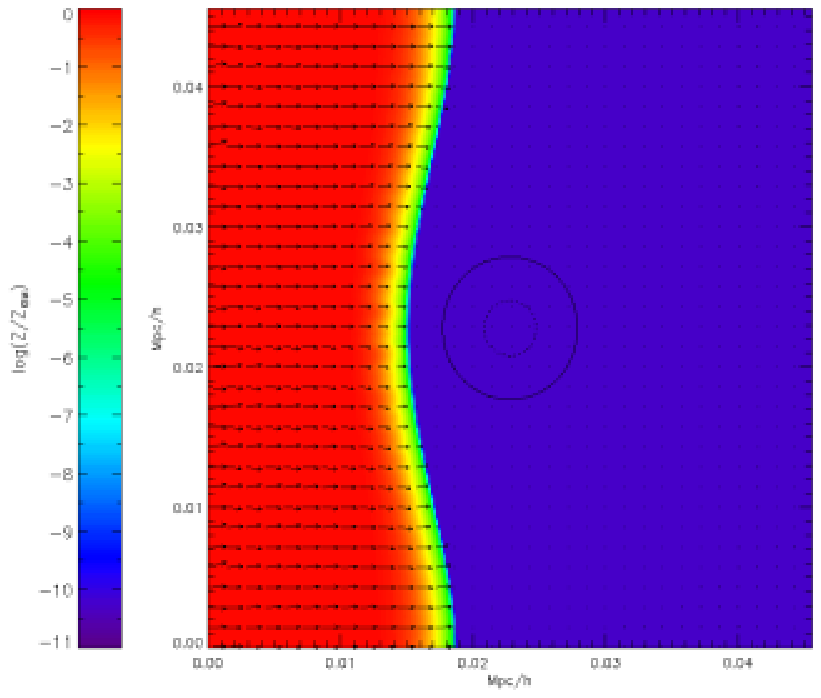}
%\hfill
%\includegraphics[width=0.43\textwidth,angle=90]{m_v30m7z6.ps}\\
\includegraphics[width=0.43\textwidth,angle=90]{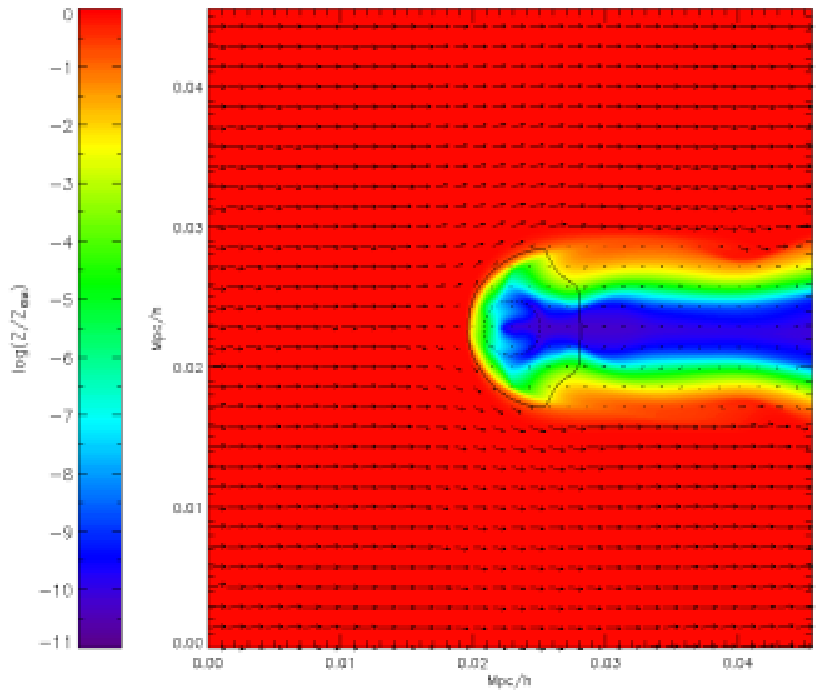}\\
\caption{The density and metallicity in a slice through the center of the box for $V_s = 30$ km/s and $M_H = 10^7 M_{\odot}$. The upper and bottom plots show density and metallicity, respectively. The left plots correspond to $z=9$ and the right ones to $z=6$. The velocity field along with contours of $\rho = \rho_{vir}$ (solid line) and $\rho = 500\rho_b$ (dotted line) are depicted in all the plots.}
\label{fig:halo1}
\end{figure}
\begin{figure}[htp]
\centering
\includegraphics[width=0.43\textwidth,angle=90]{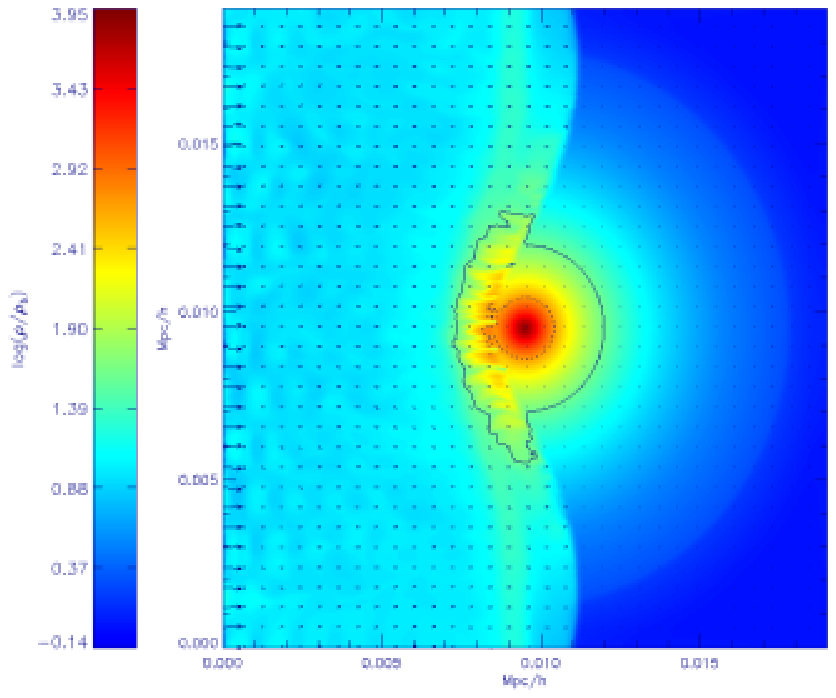}
%\hfill
%\includegraphics[width=0.43\textwidth,angle=90]{d_v100m6z6.ps}\\
%\includegraphics[width=0.43\textwidth,angle=90]{m_v100m6z9.ps}
\includegraphics[width=0.43\textwidth,angle=90]{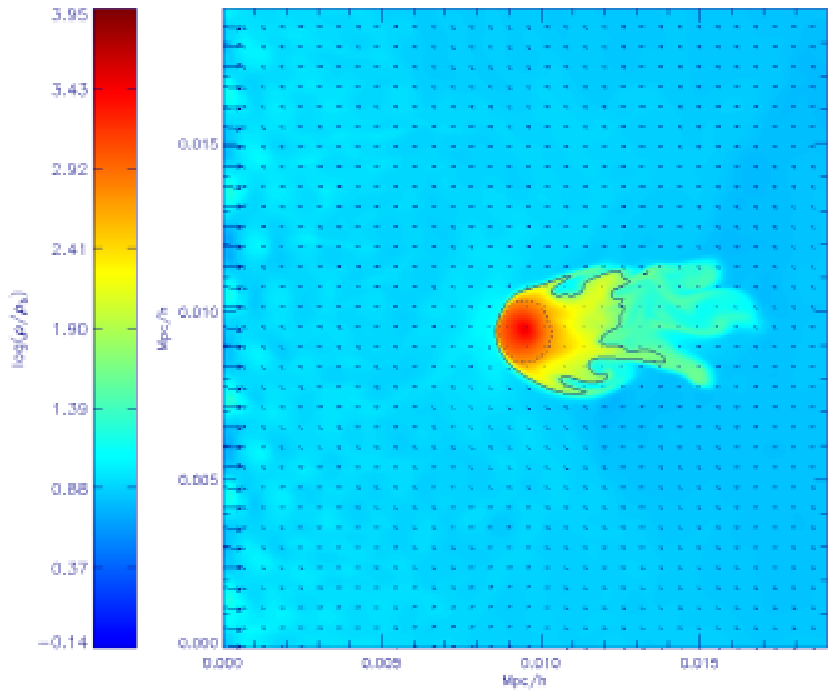}\\
\includegraphics[width=0.43\textwidth,angle=90]{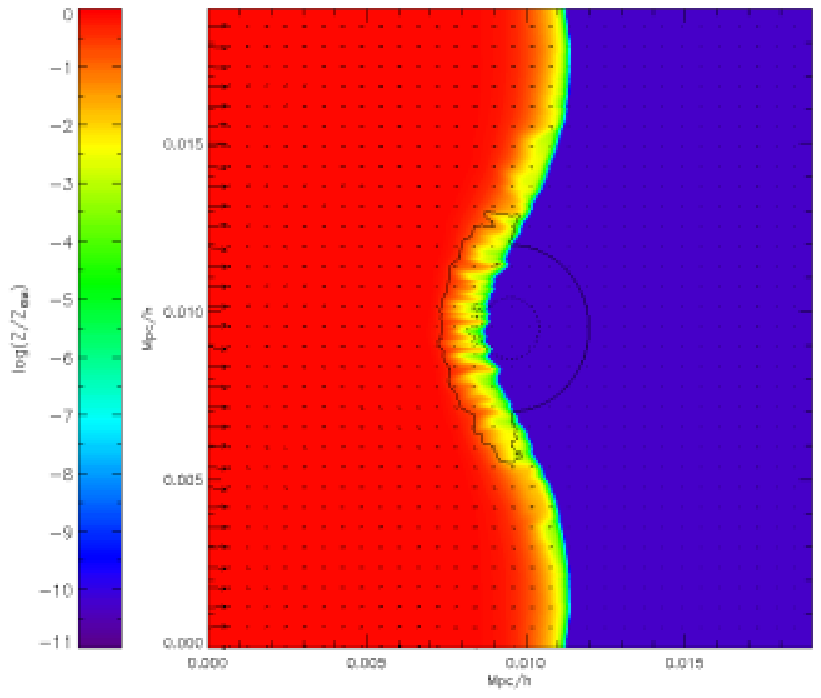}
%\hfill
%\includegraphics[width=0.43\textwidth,angle=90]{m_v100m6z6.ps}\\
\includegraphics[width=0.43\textwidth,angle=90]{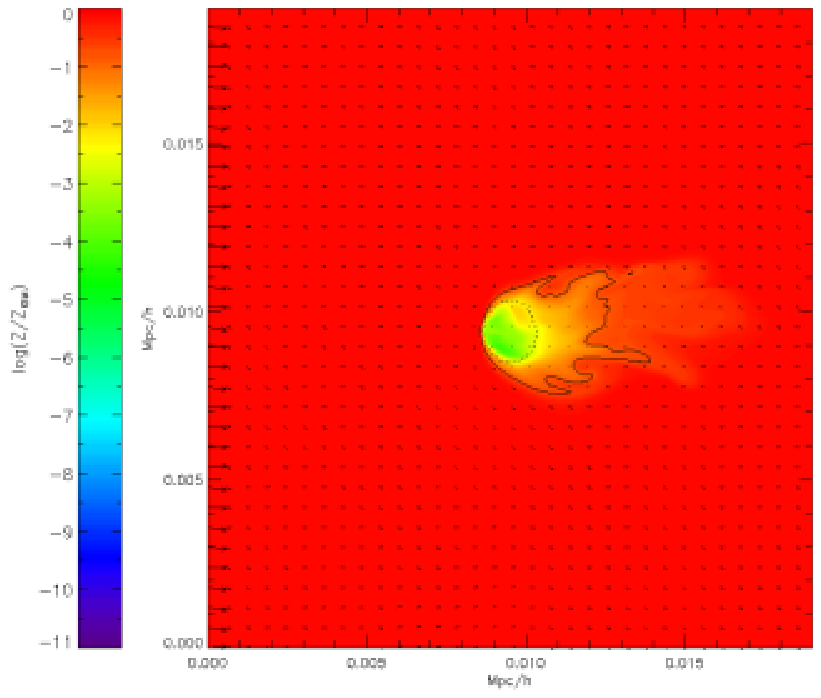}\\
\caption{The density and metallicity on a slice through the center of the box for $V_s = 100$ km/s and $M_H = 10^6 M_{\odot}$. The upper and bottom plots show density and metallicity, respectively. The left plots correspond to $z=9$ and the right ones to $z=6$. The velocity field along with contours of $\rho = \rho_{vir}$ (solid line) and $\rho = 500\rho_b$ (dotted line) are depicted in all the plots.}
\label{fig:halo2}
\end{figure}

Figures (\ref{fig:halo1}) and (\ref{fig:halo2}) show density, metallicity and velocity of the gas on a slice through the center of the halo that is perpendicular 
to the shockwave front, at $z=9$ and $z=6$.
Perhaps the most noticeable is that the gas cloud inside the minihalo
is able to 
withstand significant shockwaves and to reside 
inside the halo gravitational potential well for an extended period of time.
The gravitational potential well of the dark matter halo 
is more ``steady" than a pure self-gravitating gas cloud, 
since the dark matter is unaffected by gasdynamic processes
at the zero-th order, in agreement with M93.

\begin{figure}
%\plottwo{convergence500.ps}{convergencevir.ps}
\plottwo{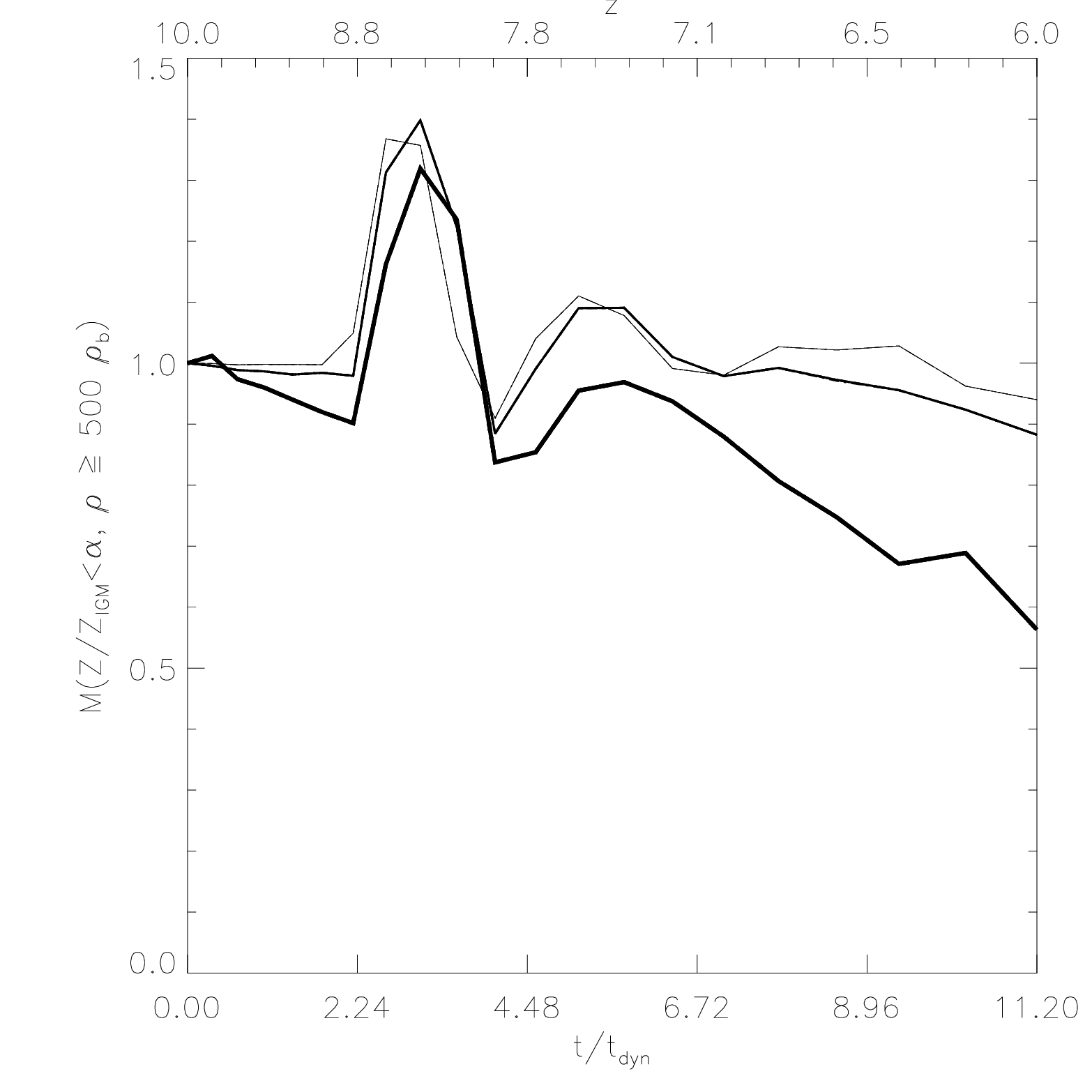}{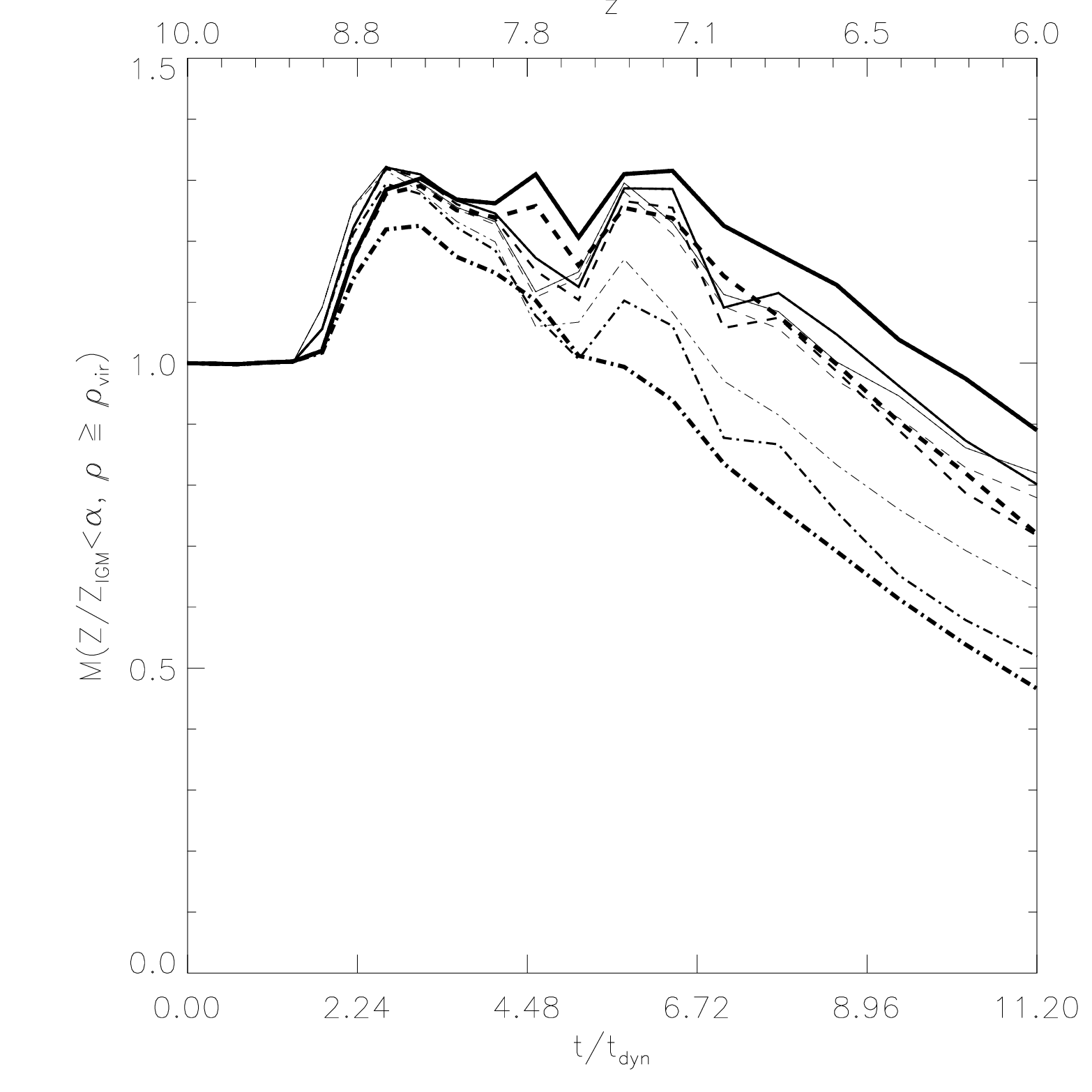}
\caption{Simulations with different resolutions were run to test convergence. Results 
of the evolution of the mass of gas in different ranges of density and metallicity using
$128^3$, $256^3$, and $512^3$ cells are presented in thick, intermediate, and thin lines, respectively.
In the three cases a $10^7 M_{\odot}$ halo with a shock velocity of 100 km/sec was simulated.
The left plot takes density range $\rho > 500 \rho_b$, where $\rho_b$ is the mean baryon 
density of the universe at z=10. The right plot takes density range $\rho > \rho_{vir}$, 
where $\rho_{vir} \approx 49.3 \rho_b$ is the density at the virial radius of the halo. 
The metallicities considered are $Z/Z_{IGM} < \alpha$, where $\alpha = 1,0.3,0.1,0.03$ 
are represented by solid, dotted, dashed, and dot-dashed lines, respectively.  The dynamical time is 
$t_{dyn} = (800\pi G\Omega_m\rho_{crit} )^{-\frac{1}{2}}$. 
}
\label{fig:converga}
\end{figure}

However, the mixing between the primordial minihalo gas and metal-enriched IGM  
due to hydrodynamical instabilities is apparent.
First, as seen in the left panels in Figure (\ref{fig:halo2}), 
the Richtmyer-Meshkov instability seems most apparent 
when the interface between the sweeping IGM and the minihalo gas cloud
is first being accelerated by the shock moving from left to right.
Subsequently, with the build-up of a smoother and larger density transition region
on the left side of the halo gas cloud and reduced shock strengths,
the Richtmyer-Meshkov instability progressively abates.
Second, as can be seen in the right panels in Figures (\ref{fig:halo1}) 
and (\ref{fig:halo2}), the Kelvin-Helmholtz instability provides 
an efficient mechanism to mix gas in the shearing regions 
at the outer part of the minihalos. 
The fact that the density peak largely remains at the center
of the dark matter halo over the extended period of time while
the outer layers become mixed with the IGM suggests that
mixing due to hydrodynamic instabilities plays the dominant role,
whereas ram-pressure stripping is sub-dominant, at least for these two cases
and during the displayed time interval.
Nevertheless, the central regions of the minihalo gas clouds
are significantly contaminated with metals at later times
(right panels in Figures \ref{fig:halo1}, \ref{fig:halo2}).
We will later show in Figure (\ref{fig:converga}) our convergence test of the results,
suggesting that our numerical resolution
appears to be adequate to properly simulate hydrodynamic
instabilities involved.

\begin{figure}
%\plottwo{500_10.ps}{vir_10.ps}
\plottwo{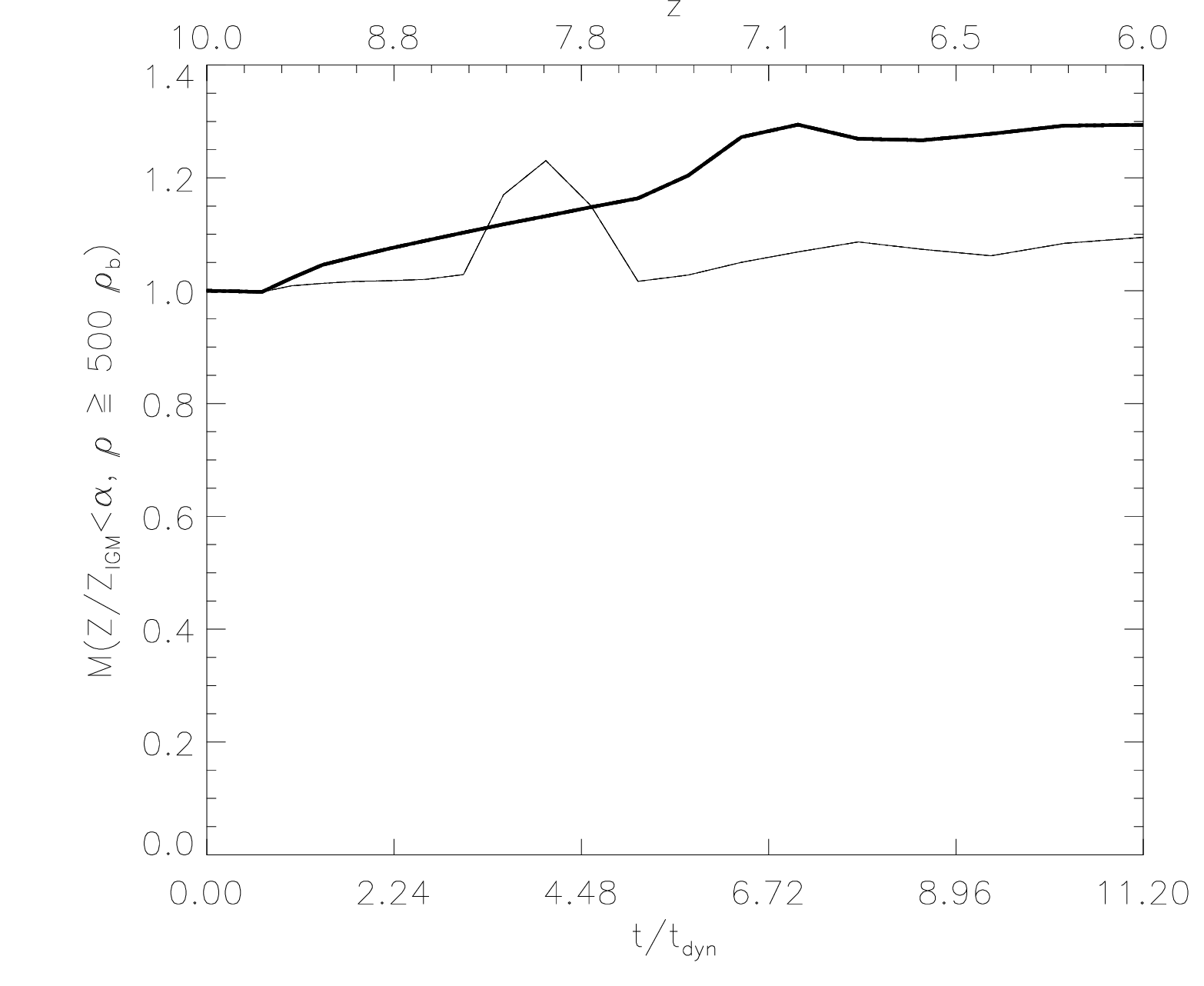}{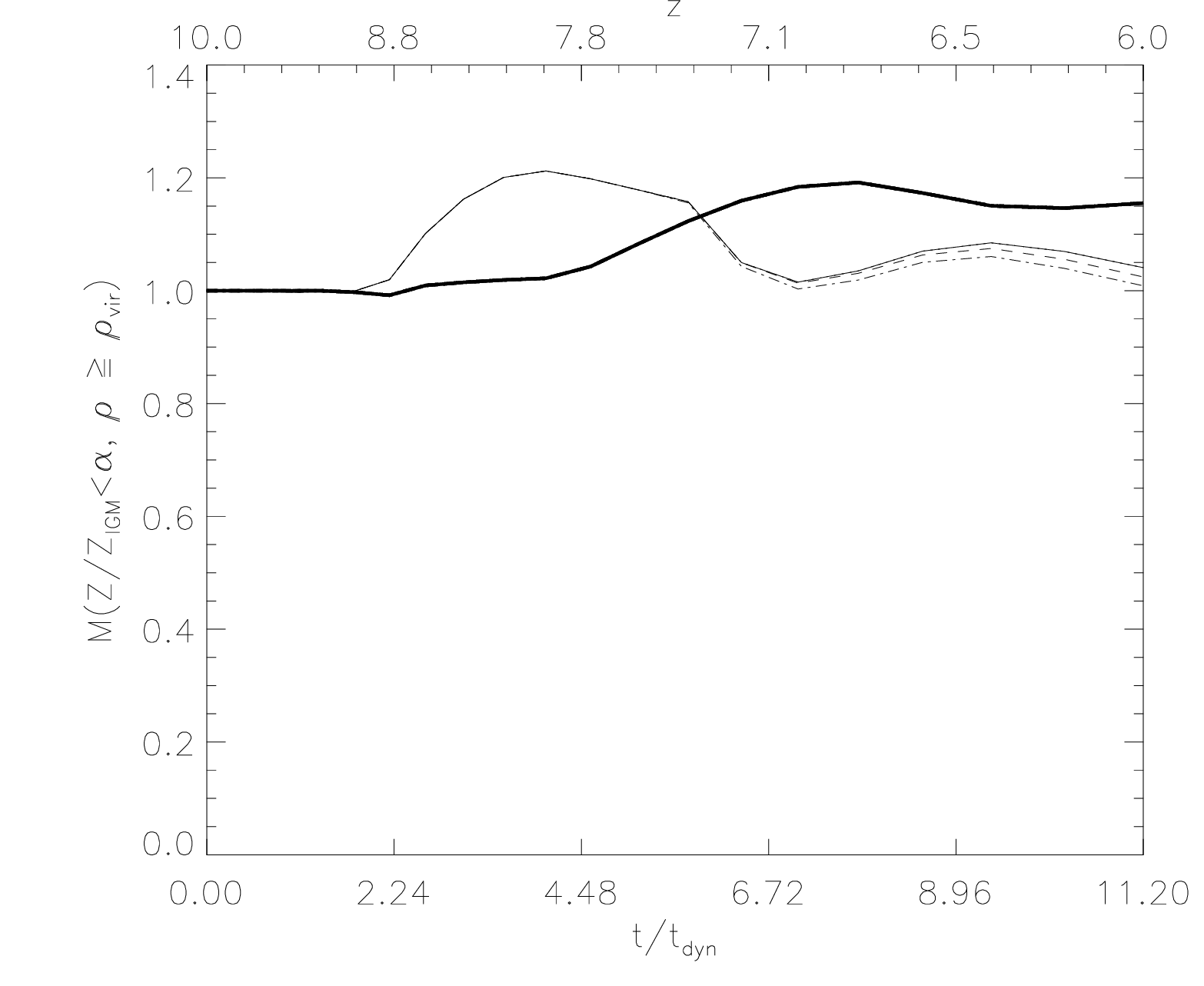}
\caption{The evolution of the mass of gas in different ranges of density and metallicity for a 
velocity shock of 10 km/sec. The masses considered are $10^6$ and $10^7 M_{\odot}$, represented 
by thin and thick lines, respectively. The left plot takes density range $\rho > 500 \rho_b$, 
where $\rho_b$ is the mean baryon density of the universe at z=10. The right plot takes 
density range $\rho > \rho_{vir}$, where $\rho_{vir}$ is the density at the virial radius of 
the halo ($\approx 40.3 \rho_b$ and $49.3 \rho_b$ for a $10^6$ and $10^7 M_{\odot}$ halos, respectively).
The metallicity ranges considered are $Z/Z_{IGM} < \alpha$, where $\alpha = 1,0.3,0.1,0.03$ 
are represented by solid, dotted, dashed, and dot-dashed lines, respectively. The dynamical time is 
$t_{dyn} = (800\pi G\Omega_m\rho_{crit} )^{-\frac{1}{2}}$.
}
\label{fig:v10}
\end{figure}

\begin{figure}
%\plottwo{500_30.ps}{vir_30.ps}
\plottwo{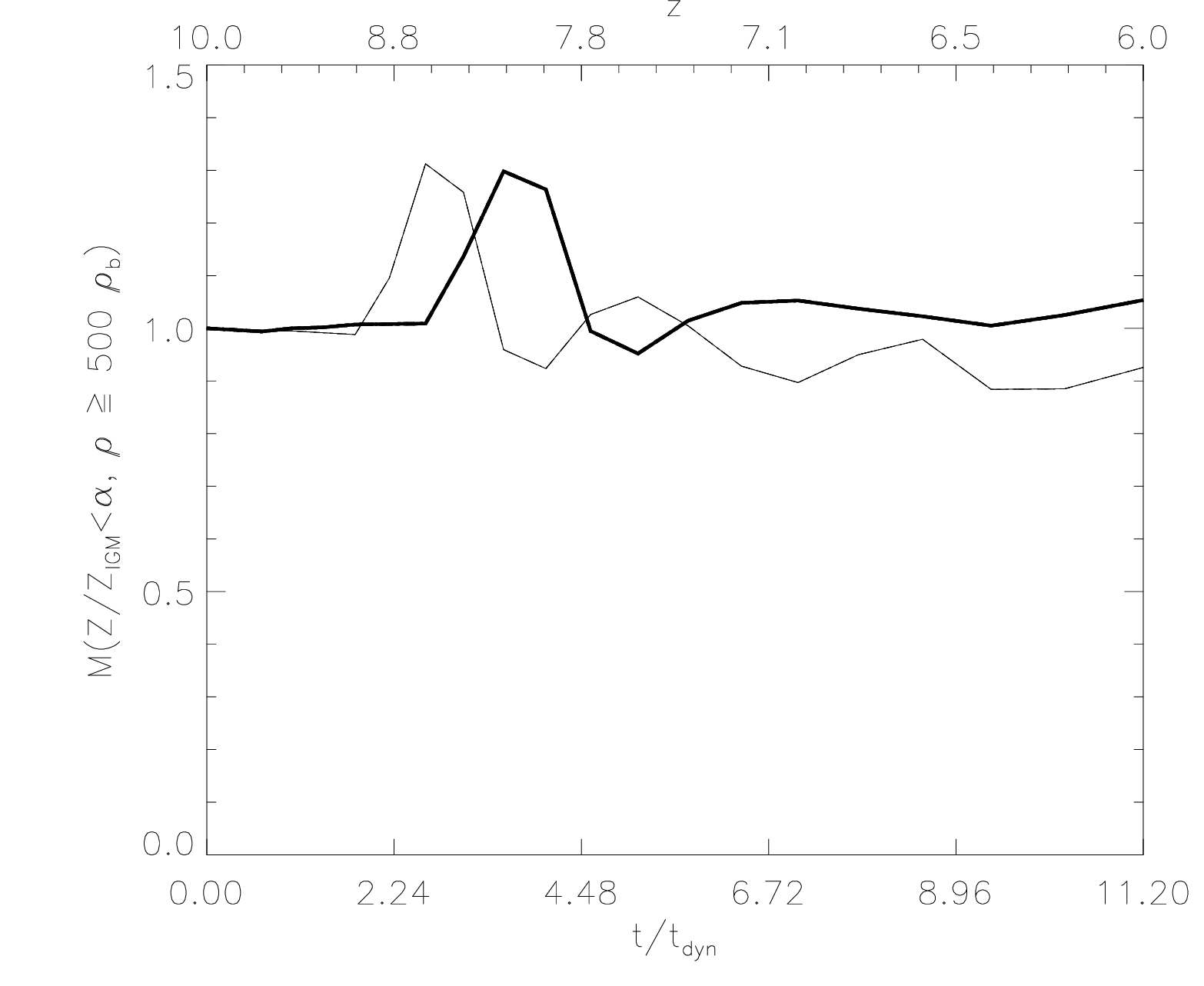}{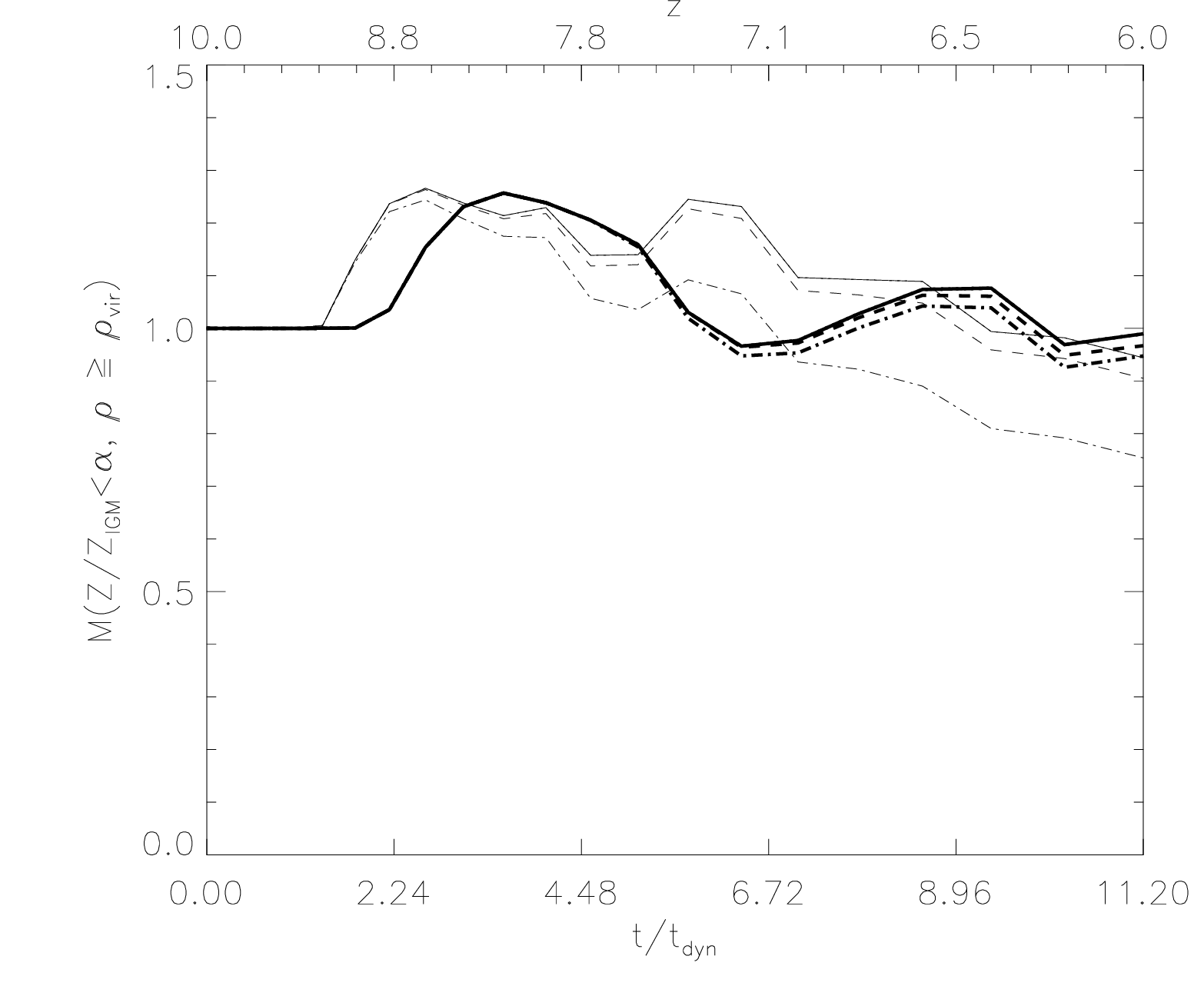}
\caption{The evolution of the mass of gas in different ranges of density and metallicity for a 
velocity shock of 30 km/sec. The masses considered are $10^6$ and $10^7 M_{\odot}$, represented 
by thin and thick lines, respectively. The left plot takes density range $\rho > 500 \rho_b$, 
where $\rho_b$ is the mean baryon density of the universe at z=10. The right plot takes 
density range $\rho > \rho_{vir}$, where $\rho_{vir}$ is the density at the virial radius of 
the halo ($\approx 40.3 \rho_b$ and $49.3 \rho_b$ for a $10^6$ and $10^7 M_{\odot}$ halos, respectively).
The metallicity ranges considered are $Z/Z_{IGM} < \alpha$, where $\alpha = 1,0.3,0.1,0.03$ 
are represented by solid, dotted, dashed, and dot-dashed lines, respectively. The dynamical time is 
$t_{dyn} = (800\pi G\Omega_m\rho_{crit} )^{-\frac{1}{2}}$.
}
\label{fig:v30}
\end{figure}

\begin{figure}
%\plottwo{500_100.ps}{vir_100.ps}
\plottwo{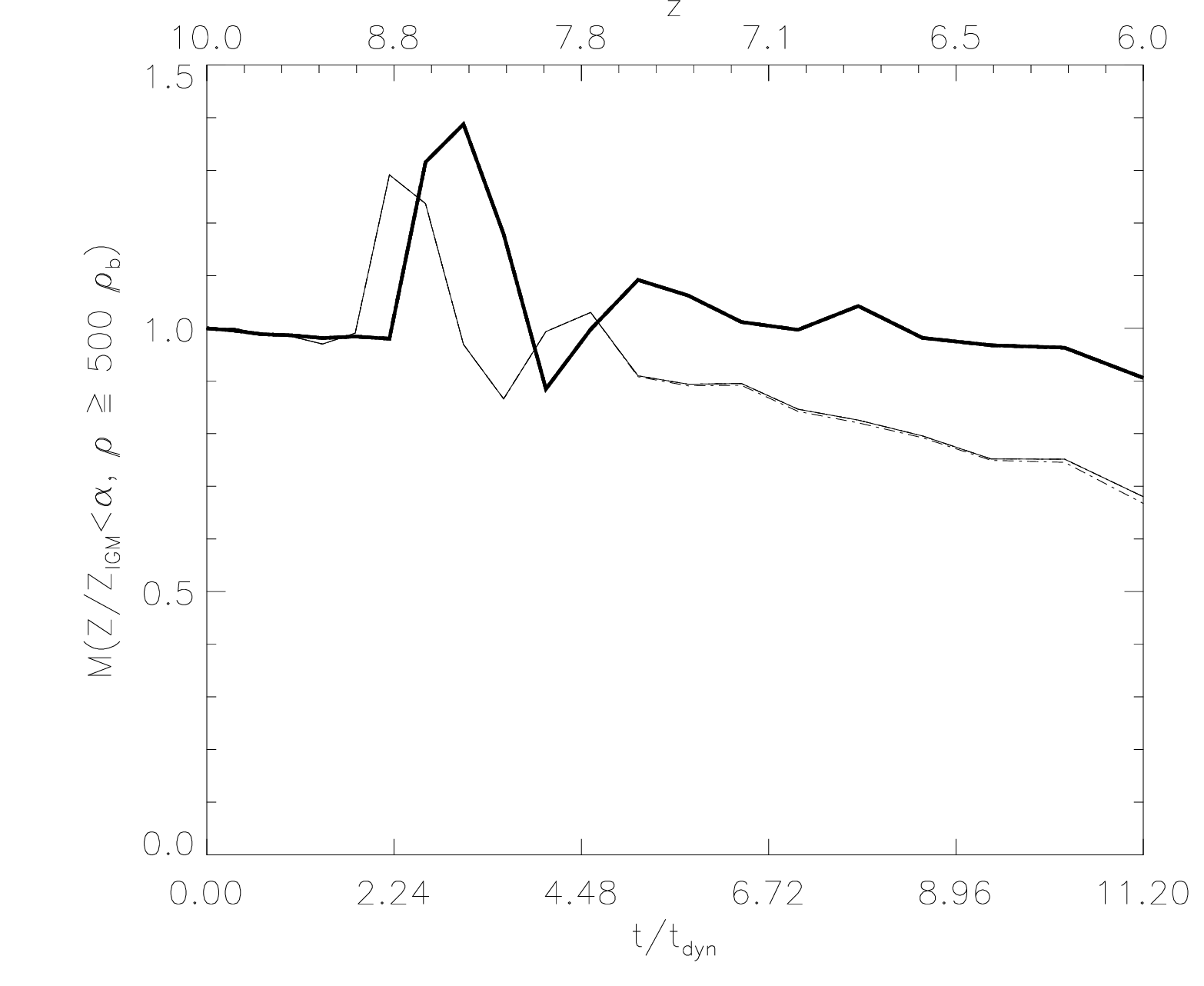}{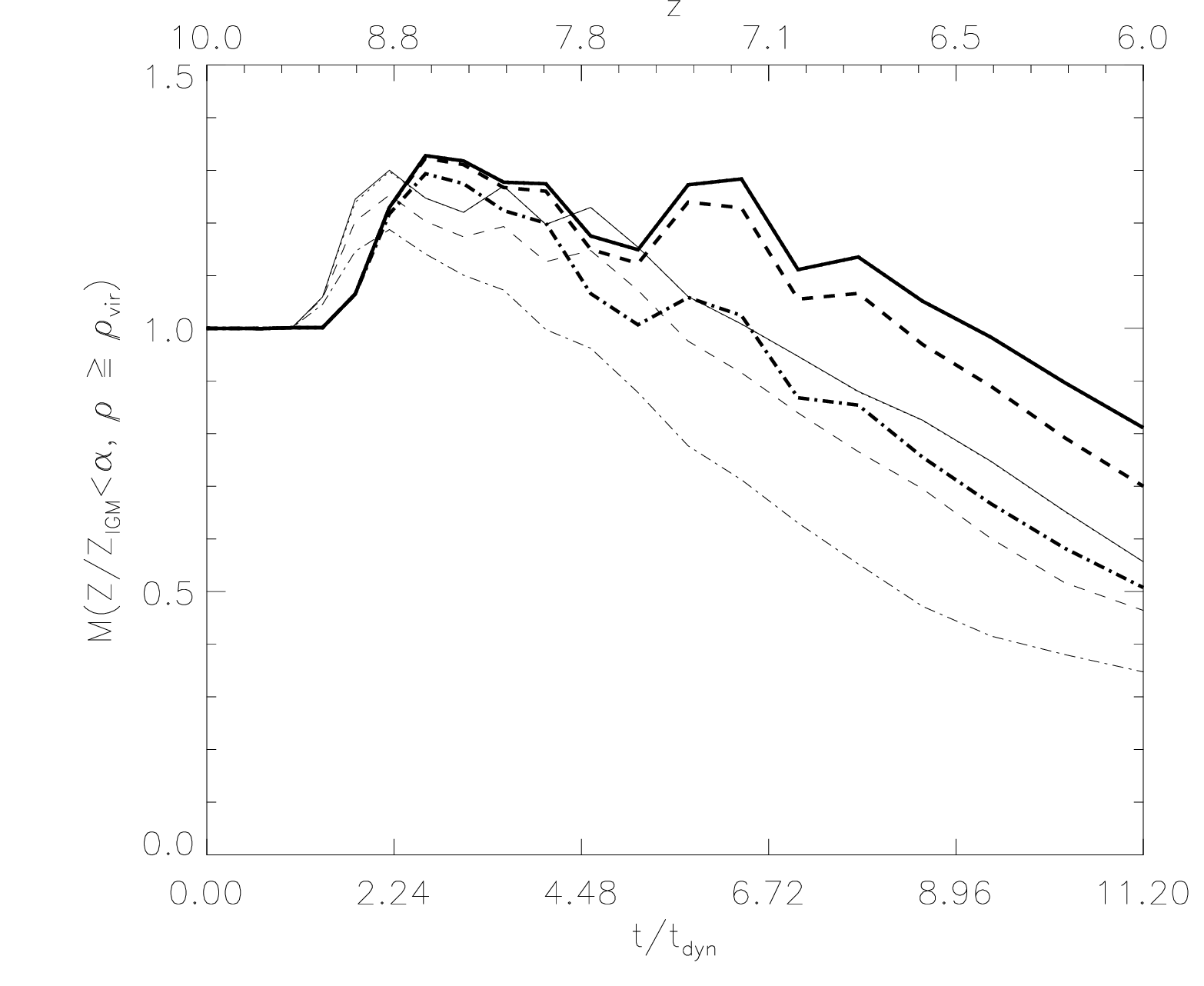}
\caption{The evolution of the mass of gas in different ranges of density and metallicity for a 
velocity shock of 100 km/sec. The masses considered are $10^6$ and $10^7 M_{\odot}$, represented 
by thin and thick lines, respectively. The left plot takes density range $\rho > 500 \rho_b$, 
where $\rho_b$ is the mean baryon density of the universe at z=10. The right plot takes 
density range $\rho > \rho_{vir}$, where $\rho_{vir}$ is the density at the virial radius of 
the halo ($\approx 40.3 \rho_b$ and $49.3 \rho_b$ for a $10^6$ and $10^7 M_{\odot}$ halos, respectively).
The metallicity ranges considered are $Z/Z_{IGM} < \alpha$, where $\alpha = 1,0.3,0.1,0.03$ 
are represented by solid, dotted, dashed, and dot-dashed lines, respectively. The dynamical time is 
$t_{dyn} = (800\pi G\Omega_m\rho_{crit} )^{-\frac{1}{2}}$.
}
\label{fig:v100}
\end{figure}

\begin{figure}
%\plottwo{500_300.ps}{vir_300.ps}
\plottwo{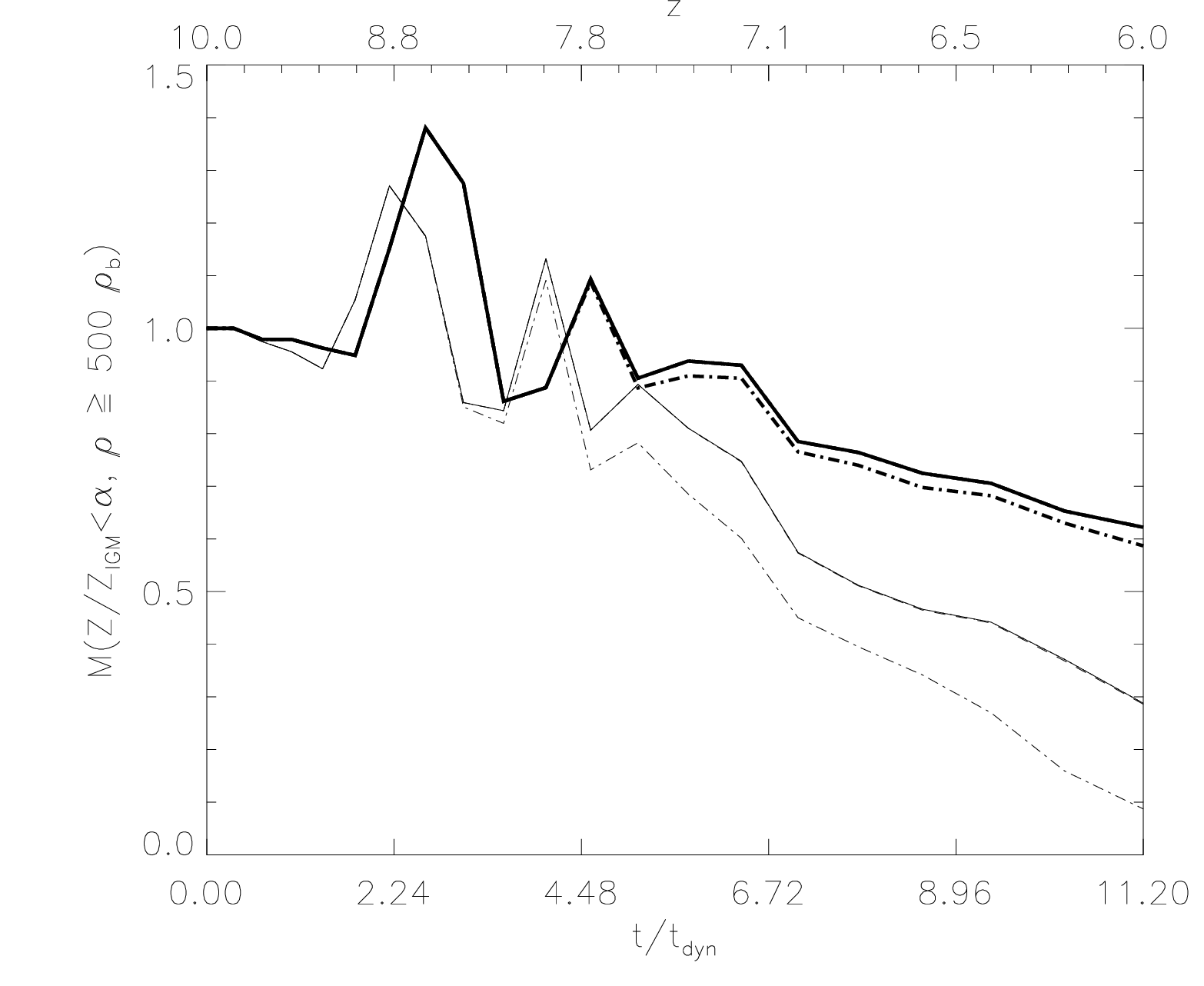}{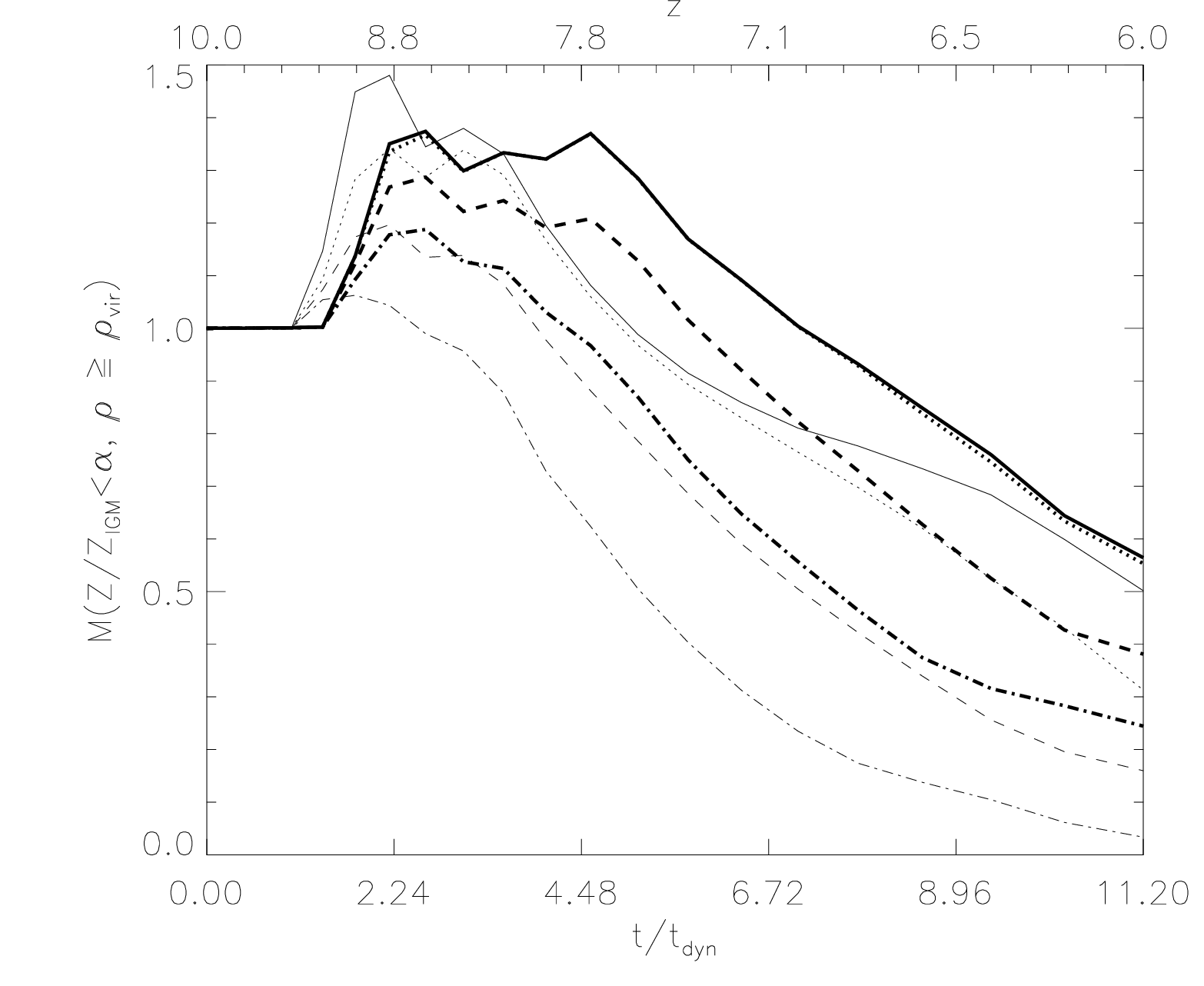}
\caption{The evolution of the mass of gas in different ranges of density and metallicity for a 
velocity shock of 300 km/sec. The masses considered are $10^6$ and $10^7 M_{\odot}$, represented 
by thin and thick lines, respectively. The left plot takes density range $\rho > 500 \rho_b$, 
where $\rho_b$ is the mean baryon density of the universe at z=10. The right plot takes 
density range $\rho > \rho_{vir}$, where $\rho_{vir}$ is the density at the virial radius of 
the halo ($\approx 40.3 \rho_b$ and $49.3 \rho_b$ for a $10^6$ and $10^7 M_{\odot}$ halos, respectively).
The metallicity ranges considered are $Z/Z_{IGM} < \alpha$, where $\alpha = 1,0.3,0.1,0.03$ 
are represented by solid, dotted, dashed, and dot-dashed lines, respectively. The dynamical time is 
$t_{dyn} = (800\pi G\Omega_m\rho_{crit} )^{-\frac{1}{2}}$.
}
\label{fig:v300}
\end{figure}

We will now turn to more quantitative results, focusing on the metal enrichment of 
gas inside minihalos by the IGM shocks.
Figures (\ref{fig:v10}-\ref{fig:v300}) show the evolution of the amount of mass
at $\rho > 500\rho_b$ and $\rho > \rho_{vir}$ respectively, 
in units of its corresponding value at the beginning of the simulation at $z=10$,
that is metal-enriched to various levels 
with $Z<\alpha Z_{IGM}$ with $\alpha=(1, 0.3, 0.1, 0.03)$.
Figures (\ref{fig:v10}-\ref{fig:v300}) show cases with $V_s = (10,30,100,300)$km/s.


From Figure (\ref{fig:v10}) with $V_s=10$km/s
we see that
for $M_H = 10^6 M_{\odot}$ only $~5\%$ of the gas is contaminated to 
$Z \ge 0.03 Z_{IGM}$ for $\rho \ge \rho_{vir}$, 
and for a $10^7M_{\odot}$ halo there is practically no gas with $Z\ge 0.03 Z_{IGM}$ 
in that density range after about 11 dynamic time (by $z\sim 6$). 
For $\rho > 500\rho_b$, there is no gas with $Z$ larger than $0.03 Z_{IGM}$
even for the $M_H = 10^6M_{\odot}$ halo. 
It is interesting to note that for this velocity 
the amount of gas at the two ranges of density considered actually increases instead of decreasing. This is due to the compression produced on the halo by the gas from the shock. 
We also observe the acoustic oscillations
in the amount of gas due to this compression. 
(The fact that acoustic oscillations for the $10^6 M_{\odot}$
halo start earlier is just due to the smaller simulation box size.)
Figure (\ref{fig:v30}) shows the case $V_s = 30$ km/s. Here we see that for $\rho > 500\rho_b$ again there is not gas that gets more metal-rich than $Z = 0.03 Z_{IGM}$
by $z\sim 6$. 
For $\rho \ge \rho_{vir}$, only $\sim 5\%$ of the gas ends up with $Z \ge 0.03 Z_{IGM}$ for a $10^7 M_{\odot}$ halo.
However, for $M_H = 10^6 M_{\odot}$, $\sim 5\%$ of the gas mass reaches
$Z \ge 0.1 Z_{IGM}$.

For $V_s = 100$ and 300 km/s (Figures \ref{fig:v100}, \ref{fig:v300})
the stripping of the outer parts of the halo becomes more important and we start to see that the amount of metal-free gas for the density ranges considered starts to decrease significantly for two reasons. First, the halo is losing a significant amount of its mass and, therefore, its global structure is being modified. So we observe a decrease in the total amount of mass for $\rho > 500\rho_b$ and $\rho > \rho_{vir}$. Second, this stripping put into contact the IGM gas with the innermost part of the halo, 
moving the mixing layer inward and increasing the efficiency of the mixing to 
higher density regions in the halo.
For $V_s = 100$ km/s, 
at $\rho \ge 500 \rho_b$ there is not significant mixing but just a small overall reduction of the mass. On the other hand, for $\rho \ge \rho_{vir}$  the total decrease of mass starts to be significant reaching even $\sim 50 \%$ for $M_H=10^6 M_{\odot}$, and the amount of gas purer than $0.03 Z_{IGM}$ is only $\sim30\%$ and $\sim50\%$ of the original
counterparts at $z = 10$ for $M_H=10^6 M_{\odot}$ and $M_H=10^7 M_{\odot}$, respectively.
For $V_s = 300$ km/s,
at $\rho \ge 500 \rho_b$ we observe significant reduction of the overall mas, especially for $M_H = 10^6 M_{\odot}$ where the mass is reduced to $\sim30\%$ percent of its original value. We can see that the mixing itself does not play a very significant role at these densities, with practically no difference between the total mass and the mass of gas with Z $<$ 0.03$\textrm{Z}_{IGM}$ for $M_H = 10^7 M_{\odot}$. The same thing happens with $M_H = 10^6 M_{\odot}$, but in this case for Z $<$ 0.1$\textrm{Z}_{IGM}$. For $\rho \ge \rho_{vir}$ we see that, along with the total reduction of mass, there is substantial enrichment of the gas. The mass of gas with  Z $<$ 0.03$\textrm{Z}_{IGM}$ is only $\sim 3\%$ and $\sim 25\%$ of its value at $z=10$ for $M_H = 10^6 \textrm{ and } 10^7 M_{\odot}$, respectively. 

To provide a convenient numerical form we 
summarize in Tables (\ref{table:z7}) and (\ref{table:z6}) 
the masses of gas in the different ranges of density and metallicity for the halo masses and shock velocities considered at the relevant redshifts $z=7$ and $z=6$, respectively.

The sensitivities of the gas cloud disruption on shock velocity
may be understood in the context of instability analysis by M93.
M93 show that, when the parameter $\eta$, defined as
\begin{equation}
\eta = {g  D R_{cl} \over 2\pi V_s^2},
\end{equation}
\noindent
is above unity, the cloud is stable up to many dynamical times,
where $D$ is the density ratio of the gas cloud to the background gas,
$R_{cl}$ is the radius of the gas cloud and
$g$ is the surface gravity.
Numerically, 
\begin{equation}
\eta(r) = ({V_s\over 22 km/s})^{-2} ({M_H\over 10^6 M_\odot})^{2/3} 
({1+z\over 11}) ({M_r\over M_H})^{-4.7}
\end{equation}
\noindent
where we have assumed that the density slope near the virial radius is
$-2.4$ \citep{NavFreWhi97}; $M_r$ is the mass with radius $r$,
$z$ is redshift.  
Equation (10) suggests that for $V_s\le 25$km/s, the gas cloud
in minihalo of mass $M_H=10^6\msun$ at $z\sim 10$
is generally quite stable, in agreement with our results.
For $V_s=300$km/s and $M_H=10^6\msun$ one obtains 
$\eta=0.005$ at $z=10$ and $M_r=M_H$, suggesting that 
the outskirts of the minihalo gas cloud would be disrupted 
on the order of a dynamic time, consistent with our results.
For $V_s=100$km/s and $M_H=10^7\msun$ we find
$\eta=0.23$ at $z=10$ and $M_r=M_H$;  
M93 find that at $\eta=0.25$, the gas mass loss is still relatively small
over many dynamic times, consistent with our simulations.

\cite{moetal02} show, in
simulations of propagation of supernova blastwaves from
$10^8h^{-1}M_\odot$ galaxies at $z=9$, that after more than a hundred
million years the relative filling factor for regions being swept by
shocks of velocities larger than $U=(10,30,100)$km/s is roughly
$(100\%, 35\%, 10\%)$.  We expect the velocities to be still smaller
at the higher redshifts of concern here, due to enhanced
cooling and larger Hubble velocity.  
Therefore, in real cosmological setting, 
combined with our findings,
we expect that a large fraction of the gas already virialized with minihalos will
be largely unaffected by metal-carrying blastwaves and remain
metal-free to modern redshift, possibly as low as $z=5-6$, 
when gas in minihalos may be photo-evaporated globally.

It is prudent to check
the convergence of computed results.
We have performed additional simulations with $128^3$ and $512^3$ grid points. 
We show in Figure (\ref{fig:converga}) an example of these convergence tests we have done. 
We see that, while the difference 
between the $128^3$ and $256^3$ cases can amount up to tens of percent
at late times (say, $t/t_{dyn}>5$),
the difference between the $256^3$ and $512^3$ cases 
is dramatically reduced  and is at a level 
of a few percent even at very late times ($t/t_{dyn}>10$).
It is instructive to notice that the tendency is to decrease the level of mixing 
as we increase the resolution. So our results must be interpreted as an upper 
limit in the metal enrichment of minihalos by shockwaves, with an accuracy of 
a few percent.

\begin{table}
\caption{Mass of gas in different ranges of density and metallicity at t = 7.07 t$_{dyn}$ (z = 7), where $t_{dyn} = (800\pi G\Omega_m\rho_{crit} )^{-\frac{1}{2}}$. Results are presented for the two kind of halos ($10^6$ and $10^7 M_{\odot}$) and the four shock velocities ($V_s =$ 10, 30, 100, and  300 km/sec) considered. The density ranges considered are $\rho > 500 \rho_b$, where $\rho_b$ is the mean baryon density of the universe at z=10 and $\rho > \rho_{vir}$, where $\rho_{vir}$ is the density at the virial radius of the halo ($\approx 40.3 \rho_b \textrm{ and } 49.3 \rho_b$ for the $10^6$ and $10^7 M_{\odot}$ halos, respectively). The mass of gas is measured in units of the total mass at a given range of density when t = 0 (z=10). The metallicity ranges considered are $Z/Z_{IGM} <$ 0.3, 0.1, and 0.03.}
\label{table:z7}
\begin{center}
\begin{tabular}{|l|l|c|c||c|c|}%|c|c||c|c|}
\hline
\multicolumn{2}{|c|}{$z=7$}&\multicolumn{2}{c||}{$M_H = 10^6 M_{\odot}$}&\multicolumn{2}{c|}{$M_H = 10^7 M_{\odot}$}\\
\cline{3-6}
\multicolumn{2}{|c|}{$t = 7.07 \textrm{ } t_{dyn}$}& $\rho>\rho_{500}$ & $\rho>\rho_{vir}$ & $\rho>\rho_{500}$ & $\rho>\rho_{vir}$ \\%& $\rho>\rho_{500}$ & $\rho>\rho_{vir}$ & 
\hline\hline
&$Z/Z_{IGM}<0.3$ &1.069&1.015&1.294&1.184\\
\cline{2-6}
$V_s = 10 \textrm{km/sec}$&$Z/Z_{IGM}<0.1$ &1.069&1.013&1.294&1.184\\
\cline{2-6}
&$Z/Z_{IGM}<0.03$ &1.069&1.003&1.294&1.184\\
\hline
\hline
&$Z/Z_{IGM}<0.3$ &0.897&1.096&1.053&0.977\\
\cline{2-6}
$V_s = 30 \textrm{km/sec}$&$Z/Z_{IGM}<0.1$ &0.897&1.072&1.053&0.972\\
\cline{2-6}
&$Z/Z_{IGM}<0.03$ &0.897&0.936&1.053&0.953\\
\hline
\hline
&$Z/Z_{IGM}<0.3$ &0.846&0.947&0.997&1.111\\
\cline{2-6}
$V_s = 100 \textrm{km/sec}$&$Z/Z_{IGM}<0.1$ &0.846&0.839&0.997&1.055\\
\cline{2-6}
&$Z/Z_{IGM}<0.03$ &0.842&0.630&0.997&0.868\\
\hline
\hline
&$Z/Z_{IGM}<0.3$ &0.574&0.765&0.785&1.003\\
\cline{2-6}
$V_s = 300 \textrm{km/sec}$&$Z/Z_{IGM}<0.1$ &0.573&0.505&0.785&0.823\\
\cline{2-6}
&$Z/Z_{IGM}<0.03$ &0.450&0.235&0.765&0.557\\
\hline
\end{tabular}
\end{center}
\end{table}
\begin{table}
\caption{Mass of gas in different ranges of density and metallicity at t = 11.2 t$_{dyn}$ (z = 6), where $t_{dyn} = (800\pi G\Omega_m\rho_{crit} )^{-\frac{1}{2}}$. Results are presented for the two kind of halos ($10^6$ and $10^7 M_{\odot}$) and the four shock velocities ($V_s =$ 10, 30, 100, and  300 km/sec) considered. The density ranges considered are $\rho > 500 \rho_b$, where $\rho_b$ is the mean baryon density of the universe at z=10 and $\rho > \rho_{vir}$, where $\rho_{vir}$ is the density at the virial radius of the halo ($\approx 40.3 \rho_b \textrm{ and } 49.3 \rho_b$ for the $10^6$ and $10^7 M_{\odot}$ halos, respectively). The mass of gas is measured in units of the total mass at a given range of density when t = 0 (z=10). The metallicity ranges considered are $Z/Z_{IGM} <$ 0.3, 0.1, and 0.03.}
\label{table:z6}
\begin{center}
\begin{tabular}{|l|l|c|c||c|c|}%|c|c||c|c|}
\hline
\multicolumn{2}{|c|}{$z=6$}&\multicolumn{2}{c||}{$M_H = 10^6 M_{\odot}$}&\multicolumn{2}{c|}{$M_H = 10^7 M_{\odot}$}\\
\cline{3-6}
\multicolumn{2}{|c|}{$t = 11.2 \textrm{ } t_{dyn}$}& $\rho>\rho_{500}$ & $\rho>\rho_{vir}$ & $\rho>\rho_{500}$ & $\rho>\rho_{vir}$ \\ 
\hline\hline
&$Z/Z_{IGM} < 0.3$ &1.095&1.041&1.294&1.155\\
\cline{2-6}
$V_s = 10 \textrm{km/sec}$&$Z/Z_{IGM} < 0.1$ &1.095&1.024&1.294&1.155\\
\cline{2-6}
&$Z/Z_{IGM} < 0.03$ &1.095&1.009&1.294&1.155\\
\hline
\hline
&$Z/Z_{IGM} < 0.3$ &0.926&0.944&1.054&0.990\\
\cline{2-6}
$V_s = 30 \textrm{km/sec}$&$Z/Z_{IGM} < 0.1$ &0.926&0.905&1.054&0.967\\
\cline{2-6}
&$Z/Z_{IGM} < 0.03$ &0.926&0.754&1.054&0.948\\
\hline
\hline
&$Z/Z_{IGM} < 0.3$ &0.680&0.556&0.906&0.811\\
\cline{2-6}
$V_s = 100 \textrm{km/sec}$&$Z/Z_{IGM} < 0.1$ &0.680&0.464&0.906&0.700\\
\cline{2-6}
&$Z/Z_{IGM} < 0.03$ &0.667&0.347&0.906&0.508\\
\hline
\hline
&$Z/Z_{IGM} < 0.3$ &0.288&0.313&0.622&0.553\\
\cline{2-6}
$V_s = 300 \textrm{km/sec}$&$Z/Z_{IGM} < 0.1$ &0.286&0.160&0.622&0.381\\
\cline{2-6}
&$Z/Z_{IGM} < 0.03$ &0.087&0.034&0.587&0.244\\
\hline
\end{tabular}
\end{center}
\end{table}
\section{Conclusions}
\label{Conclusiones}

It is frequently assumed that the metallicity
of the intergalactic medium is the primary determinant
of the epoch of the transition
from Pop-III to Pop-II stars.
We wish to point out a potentially large difference
between the metallicity of the intergalactic medium and the metallicity
of the gas in minihalos. 
Utilizing hydrodynamic simulations of gas clouds in minihalos
subject to destructive processes associated with  
the encompassing intergalactic shocks carrying metal-enriched gas,
we find that a large fraction of gas in virialized minihalos
remains at a metallicity much lower than that of the intergalactic medium.
For example, for realistic shocks of velocities of $10-100$km/s,
more than ($90\%,65\%$) of the high density gas with $\rho \ge 500\rho_b$
inside a minihalo virialized at $z=10$ of mass $(10^7,10^6)\msun$
remains at a metallicity lower than 3\% of that of the intergalactic medium
by redshift $z=6$,
under the harsh condition that the minihalo is exposed to 
shockwaves continuously from $z=10$ to $z=6$.

In the standard cosmological model, if large halos with efficient atomic
cooling are responsible for producing most of the reionizing photons,
smaller minihalos virialize before the universe
is significantly reionized.
Thus, gas in virialized minihalos may provide 
an abundant reservoir of primordial gas to possibly allow for the 
formation of Population-III metal-free stars to extend to 
much lower redshift than expected otherwise based on
the enrichment of intergalactic medium.

A related issue that is not addressed here concerns
the fate of the gas inside minihalos
when exposed to reionizing photons.
The situation is complicated because the timescale of the 
photo-evaporation of gas in minihalos
\citep{baretal02,ili05,cia06} 
may be $\sim 100$Myrs \citep{sha04}; 
the timescale may be still longer at higher redshifts ($z>10$) 
and/or at lower ionizing fluxes than used in the work of \cite{sha04}.
It may be that a full understanding requires
detailed calculations that incorporate both radiative transfer 
and metal-enrichment processes.

\acknowledgements
We gratefully acknowledge financial support by
grants AST0407176, NNG06GI09G and financial support from Princeton University.

\end{document}